\documentstyle[preprint,aps,prb,eqsecnum]{revtex}

\begin{document}
\draft
\title{Magnetic impurity coupled to interacting conduction
electrons}
\author{Tom Schork}
\address{Max-Planck-Institut f\"ur Physik komplexer Systeme,\\
Bayreuther Str.\ 40 Haus 16, D-01187 Dresden, Germany}
\date{\today}
\maketitle

\begin{abstract}
We consider a magnetic impurity which interacts by hybridization
with a system of weakly correlated electrons and determine the
energy of the ground state by means of an $1/N_f$ expansion. The
correlations among the conduction electrons are described by a
Hubbard Hamiltonian and are treated to lowest order in the
interaction strength. We find that their effect on the Kondo
temperature, $T_{\rm K}$, in the Kondo limit is twofold: First,
the position of the impurity level is shifted due to the
reduction of charge fluctuations, which reduces $T_{\rm
K}$. Secondly, the bare Kondo exchange coupling is enhanced as
spin fluctuations are enlarged. In total, $T_{\rm K}$
increases. Both corrections require intermediate states beyond
the standard Varma-Yafet ansatz. This shows that the Hubbard
interaction does not just provide quasiparticles, which hybridize
with the impurity, but also renormalizes the Kondo coupling.
\end{abstract}


\section{Introduction}
\label{sec:intro}

Recently, heavy-fermion behavior has been observed in the
electron-doped cuprate Nd$_{2-x}$Ce$_x$CuO$_4$
\hbox{($0.1\lesssim x\lesssim 0.2$)}.\cite{Brugger93} Below 0.3~K
a linear specific heat \hbox{$C_v = \gamma T$} is observed with a
large Sommerfeld coefficient \hbox{$\gamma \simeq 4{\rm
J}/(\mbox{mole Nd}\cdot {\rm K^2})$}. In the same temperature
regime, the spin susceptibility is found to be independent of the
temperature and the Sommerfeld-Wilson ratio is of order
unity. These are characteristic features of heavy-fermion
excitations.\cite{Fulde88} However, the characteristic low energy
scale of the order of 1~K which is associated with this behavior
cannot be explained by applying the usual theory of the Kondo
effect which assumes that the conduction carriers behave as free
particles.\cite{Fulde93} This is not too surprising because
undoped $\rm Nd_2CuO_4$ is an antiferromagnetic charge-transfer
insulator instead of a metal,\cite{Skanthakumar89,Oseroff90}
despite of one hole per unit cell. Upon doping the Nd ions are
therefore coupling to a system of strongly correlated
electrons\cite{Boothroyd90,Adelmann92} rather than to weakly or
uncorrelated ones.

{\em Hamiltonian and Scaling.} In order to explain this new type
of heavy-fermion behavior, it has therefore been proposed to
include the correlations among the conduction electrons by
including an on-site repulsion.\cite{Fulde93,Schork94} Thus, the
total Hamiltonian
\begin{equation}
H = H_c + H_f + H_{cf}
\label{Ham}
\end{equation}
goes beyond that of the single-site Anderson impurity
model.\cite{Anderson61} $H_c$ is a Hubbard Hamiltonian describing
the conduction electrons
\begin{equation}
\begin{array}{r c l}
H_c & = & \displaystyle H_t + H_U  \phantom{\sum_{k,\sigma}}
\\
H_t & = & \displaystyle \sum_{k,\sigma} \epsilon(k)
c^\dagger_{k\sigma} c^{\phantom{\dagger}}_{k\sigma}
\\
H_U & = & \displaystyle \frac{\tilde U}{2N_s}
\sum_{kk'q,\sigma\neq\sigma'}
:c^\dagger_{k+\delta\sigma} c^{\phantom{\dagger}}_{k\sigma}
c^\dagger_{k'-\delta\sigma'} c^{\phantom{\dagger}}_{k'\sigma'}:
{}~.
\label{Ham_c}
\end{array}
\end{equation}
$c^\dagger_{k\sigma}$ creates an electron with spin $\sigma$ and
momentum $k$, $N_s$ is the number of lattice sites. The
non-interacting dispersion is given by $\epsilon(k)$. $:\cdots:$
denotes normal ordering with respect to the Fermi sea $|{\rm
FS}\rangle$ where all states below the Fermi momentum, $k_{\rm
F}$, are occupied. The magnetic impurity is assumed to contain
one orbital (e.g., 4$f$), which is either empty or singly
occupied. Double occupancies are excluded because of the strong
repulsion of electrons in that orbital. The energy of the
$f$-orbital is then given by
\begin{equation}
H_{f} = \epsilon_f \sum_{\sigma}
\hat f^\dagger_{\sigma} \hat f^{\phantom{\dagger}}_{\sigma} ~,
\end{equation}
where $\hat f^\dagger_\sigma = |\sigma\rangle \langle 0|$ are
Hubbard operators forbidding a double occupancy of the impurity
site and \hbox{$\epsilon_f<0$}. The two subsystems are coupled by
a local hybridization
\begin{equation}
\displaystyle H_{cf}
= \frac{\tilde V}{\sqrt{N_s}} \sum_{k,\sigma}
\left(
\hat f^\dagger_\sigma c^{\phantom{\dagger}}_{k\sigma}
+ c^\dagger_{k\sigma} \hat f^{\phantom{\dagger}}_\sigma
\right) ~.
\label{Ham_cf}
\end{equation}

Taking a twofold degeneracy of the $f$-orbital ($\sigma=1,2$),
the model defined in Eqs.~(\ref{Ham}--\ref{Ham_cf})
corresponds\cite{Fulde93} to the situation found in
Nd$_{2-x}$Ce$_x$CuO$_4$ since the crystal-field ground state of
Nd is a doublet.\cite{Boothroyd92} In order to perform a
systematic expansion we set $\sigma=1\dots N_f$ and consider
large $N_f$. This generalization deserves some comment: If the
conduction electrons are uncorrelated ($\tilde U=0$), this
corresponds to treating an $N_f$-fold degenerate impurity which
hybridizes with an $s$-wave conduction band. This is seen by
expanding the conduction electron states in partial waves about
the impurtity site and, assuming a spherically symmetric
hybridization, only conduction electrons with the same total
angular momentum are coupled to the impurity while the others
play a passive role and can be dropped.\cite{HewsonBook}

Due to the interactions among the conduction electrons this
change of basis does not simplify the Hamiltonian~(\ref{Ham}).
Nevertheless, we will consider the Hamiltonian in
Eqs.~(\ref{Ham}--\ref{Ham_cf}) for $\sigma = 1\dots N_f$, which
could be viewed as an SU($N_f$) generalization of the original
model which has SU(2) symmetry. We thereby create an artificial
model, which no longer corresponds to the physical situation of
an $N_f$-fold degenerate impurity hybridizing with a correlated
$s$-band. The advantage in doing so is, however, that a
controlled approximation becomes possible for this model, namely
an expansion in $1/N_f$.\cite{lattice}

In taking the limit $N_f\to\infty$, we keep the density of
conduction electrons per spin constant, so that the kinetic
energy increases $\propto N_f$. To have a proper limit
$N_f\to\infty$, the hybridization coupling constant $\tilde V$
has to be scaled according to $\tilde V =
V/\sqrt{N_f}$.\cite{HewsonBook} As regards the Hubbard
interaction, we set $\tilde U/2= U/(2N_f)$ as was suggested in
Ref.~\onlinecite{Marston89}. With this scaling, the correction to
the ground-state energy of the Hubbard model~(\ref{Ham_c}) is of
order $N_f^0$, both in second-order perturbation theory in $U$
and when summing the diagrams of the random-phase approximation
(RPA), which is one order less than the $U=0$ energy.

{\em A straightforward variational ansatz.} In case the
conduction electrons are uncorrelated ($U=0$) Varma and
Yafet\cite{Varma76} proposed the following variational ansatz for
the ground-state wave function of~(\ref{Ham})
\begin{equation}
|\Psi_{0}\rangle =
\left( 1 +
\sum_{q\sigma} \alpha_q
\hat f^\dagger_\sigma c^{\phantom{\dagger}}_{q\sigma}
\right)
|{\rm FS}\rangle ~.
\label{VY_orig}
\end{equation}
$|{\rm FS}\rangle$ denotes the filled Fermi sea with empty
$f$-level. Via $H_{cf}$, it couples to the states
$f^\dagger_\sigma c^{\phantom{\dagger}}_{q\sigma} |{\rm
FS}\rangle$. Each of them describes a singlet formed between the
$f$ level and the free electron state with momentum $q$, where
$q$ is restricted to occupied states ($|q|\leq k_{\rm
F}$). Minimizing $\langle\Psi_0 |H-E_{\rm S}|\Psi_0\rangle$ with
respect to $\alpha_q$ yields an approximate ground-state energy
$E_{\rm S}$. Compared to the energy $E_{\rm M}$ of the multiplet
$f^\dagger_\sigma |{\rm FS}\rangle$, this collective singlet
formation gives rise to a gain in kinetic energy. With this
energy gain a characteristic temperature scale $T_{\rm K}$, the
Kondo temperature, is associated. In the Kondo limit
($|\varepsilon| \ll D \ll |\epsilon_f|$, $2D=$ band width) one
finds\cite{Varma76} (in units of the band width)
\begin{equation}
T_{\rm K} = n \exp\left( \frac{\epsilon_f-\mu}{\rho V^2}\right)
{}~.
\label{TK0}
\end{equation}
Here we assumed a constant density of states $\rho=1/(2D)$ of the
conduction electrons. $\mu$ is the chemical potential of the
conduction electrons and $n = (D+\mu) \rho$ denotes the filling
per spin. Subsequently, it has been shown that in an expansion in
the inverse degeneracy of the magnetic impurity, the
ansatz~(\ref{VY_orig}) yields the ground-state energy to order
$(1/N_f)^0$.\cite{Gunnarsson83}

In the case $U\neq 0$, it is, therefore, tempting to generalize
the ansatz~(\ref{VY_orig}) by replacing the non-interacting
ground state $|{\rm FS}\rangle$ by the (unknown) one of the
Hubbard model, $|g\rangle$.\cite{Schork95} The expectation values
with respect to $|g\rangle$ which arise in a variational
calculation are given by the moments of the spectral function of
the Hubbard model, which can be taken from, e.g., applying the
projection technique.\cite{Schork95} If we assume Fermi-liquid
behavior for the Hubbard model, the result of this generalized
ansatz is obvious. We introduce quasiparticles $\tilde c^\dagger$
via $c^\dagger_{q\sigma} = \sqrt{Z} \tilde c^\dagger_{q\sigma} +
\dots$ where $Z$ denotes their renormalization factor. These
quasiparticles hybridize with the impurity site rather than bare
electrons, the effective hybridization being, however,
renormalized by $\sqrt{Z}$. Therefore, we expect a Kondo
temperature
\begin{equation}
T_{\rm K} \propto
\exp\left( \frac{\epsilon_f}{\rho_{\rm QP} Z V^2}\right) ~,
\label{TKgen}
\end{equation}
where $\rho_{\rm QP}$ is the quasiparticle density of states at
the chemical potential. Noting that $\rho = Z\rho_{\rm QP}$ is
the many-particle density of states we see that the correlations
enter only via $\rho$. In particular, $T_{\rm K}$ is not modified
for small $U$. This is in contrast to
Ref.~\onlinecite{Khaliullin94} where it has been shown for the
Kondo model by a mean-field decoupling that due to polarization
effects the Kondo temperature increases, even to lowest order in
$U$. In the strongly correlated case Eq.~(\ref{TKgen}) cannot be
correct as well since the Kondo exchange coupling should be
$V^2/U$ rather than $V^2/\epsilon_f$.\cite{Schork94}

To clarify the quality of the variational approach, we will
restrict ourselves to the weakly correlated case in this paper
and perform a $1/N_f$ expansion to lowest order in $U$. The
theoretical framework, the Brioullin-Wigner perturbation series,
is introduced in the next section. In particular, we will show
that to order $1/N_f$ additional contribution arise from the
Hubbard interaction in the singlet channel which do not occur in
the multiplet channel (Sec.~\ref{sec:ground_state}) and conclude
that these contributions modify the Kondo temperature. They are
estimated in Sec.~\ref{sec:estimate} and the results are
discussed in Sec.~\ref{sec:conclusion}.

\section{Brioullin-Wigner perturbation theory}
\label{sec:pert_theory}

The $1/N_f$-expansion for the ground-state energy can be derived
with the help of Brioullin-Wigner perturbation
theory:\cite{Gunnarsson83,Kuramoto84,Bickers87} We decompose $H$
from Eq.~(\ref{Ham}) into $H_0+H_1$ and choose $H_1 = H_U +
H_{cf}$ as perturbation. In order to obtain the singlet
ground-state energy we take as unperturbed ground state the
filled Fermi sea $|{\rm FS}\rangle$. The energy $E_{\rm S}$ of
the ground state of $H$ (relative to the energy of $|{\rm
FS}\rangle$) is given by\cite{NegeleBook}
\begin{equation}
E_{\rm S} =
\langle {\rm FS}
| H_1 \sum_{n=0}^{\infty}
\left( \frac{Q}{E_{\rm S} - \tilde H} H_1 \right)^n |
{\rm FS} \rangle ~.
\label{BW_pert}
\end{equation}
Here, $Q=1-|{\rm FS}\rangle\langle{\rm FS}|$ and $\tilde H = L_t
+ H_f$, where the Liouvillean $L_t$ is defined by $L_t A =
[H_t,A]_-$. Equation~(\ref{BW_pert}) is equivalent to the zero
temperature limit of the equation for the lowest lying pole of
the empty $f$-state propagator that appears in the partition
function (see, e.g., Ref.~\onlinecite{Bickers87}).

{\em Diagrams.}  The individual terms of the
series~(\ref{BW_pert}) can be visualized by diagrams: In $H_1$,
each $H_{cf}$ changes the occupation of the impurity level from 0
(wiggled line) to 1 (dashed line) destroying an conduction
electron (solid line), since no double occupancy is allowed (and
vice versa). This vertex carries a factor $V/\sqrt{N_f N_s}$. The
impurity line changing always between occupied and unoccupied
$f$-level constitutes the backbone of a diagram. $H_1$ contains
$H_U$ as well. The vertex $H_U$ has two incoming and two outgoing
conduction electron lines. It yields a factor $U/(N_f N_s)$ and a
$\delta$-function ensuring momentum conservation. Taking the
expectation value with respect to $|{\rm FS}\rangle$ we connect
the conduction electron lines in all possible ways. The resolvent
$Q/(\varepsilon - \tilde H)$ yields the energy of the
intermediate states and, because of the Liouvillean $L_t$ only
the energy difference with respect to the filled Fermi sea
enters. Conduction electron lines pointing to the right
correspond to particle-like excitations (with a momentum denoted
by a capital letter, $|Q| > k_{\rm F}$) while those pointing to
the left are hole-like (denoted by $|q| \leq k_{\rm F}$). Without
Hubbard interaction $H_U$ these rules correspond to the standard
ones\cite{HewsonBook,Bickers87} for the self energy of the
propagator of the empty state in the partition function at zero
temperature.

For an expansion in $1/N_f$ we note that each closed loop of
fermions yields a summation over spin and, hence, a factor $N_f$,
whereas each $V$-vertex is $\propto 1/\sqrt{N_f}$. To lowest
order in $1/N_f$, the application of $H_U$ does not change the
order of a diagram: If we connect two conduction electron lines
of a closed loop by the 4-point interaction $H_U$ we create two
loops with the only restriction $\sigma\neq\sigma'$, which is of
higher order in $1/N_f$.

{\em Multiplet energy.} Similarly to Eq.~(\ref{BW_pert}), we
obtain the energy $E_{\rm M}$ for the multiplet ground state by
taking the expectation values with respect to the multiplet state
$f^\dagger_\sigma |{\rm FS}\rangle$. The resulting equation
corresponds to the lowest pole of the propagator of the occupied
$f$-state. The important energy for the low-temperature
thermodynamics is given by the energy difference of singlet and
multiplet ground state, $E_{\rm M} - \mu - E_{\rm S}$, which is
related to the Kondo temperature.\cite{Gunnarsson85} Note that
the multiplet has one electron more than the singlet in our
definition.

{\em Renormalization of the bare propagators.}  Already to order
$(1/N_f)^0$ the bare empty $f$-level (single wiggle line, $1/z$)
has to be renormalized (double wiggle line, $G_0(z)$). This
renormalization arises from a partial summation in
Eq.~(\ref{BW_pert}) shown in Fig.~\ref{renorm}:
\begin{equation}
\displaystyle G_0(z)
= \frac{1}{z} + \frac{1}{z} I^{(0)}(z) G_0(z)
= \frac{1}{z-I^{(0)}(z)} ~.
\end{equation}
The self-energy $I^{(0)}(z)$ (see also Fig.~\ref{Singlet}{\em a})
evaluates to
\begin{equation}
\displaystyle I^{(0)}(z)
= \frac{V^2}{N_s} \sum_q \frac{1}{z + \epsilon_q - \epsilon_f} ~.
\label{self}
\end{equation}
The propagator of the occupied $f$-state, $G_1(z)$, is not
renormalized to this order
\begin{equation}
\displaystyle G_1(z) = \frac{1}{z-\epsilon_f} ~.
\end{equation}

\section{Ground-state energies to order $1/N_f$}
\label{sec:ground_state}

{\em Diagrams for the singlet energy.} To order $(1/N_f)^0$ only
the diagram shown in Fig.~\ref{Singlet}{\em a} occurs. It was
already evaluated in Eq.~(\ref{self})
\begin{equation}
I^{(0)}(z)
= \displaystyle \frac{V^2}{N_s} \sum_q G_1(z+\epsilon_q) ~.
\label{sum_S0}
\end{equation}
There are no diagrams $\propto U$ to this order. To order $1/N_f$
we first find the diagram shown in Fig.~\ref{Singlet}{\em b}:
\begin{equation}
\displaystyle I^{(1)}(z)
= \frac{V^4}{N_f N_s^2}
\sum_{qQ} \left[ G_1(z+\epsilon_q) \right]^2
G_0(z+\epsilon_q-\epsilon_Q) ~.
\label{sum_S10}
\end{equation}

As mentioned previously, applying $H_U$ does not change the order
of a diagram to lowest order in $1/N_f$. Therefore, $I^{(1)}$ can
be regarded as parent diagram in which we insert vertices of the
interaction, $H_U$. Thereby we restrict ourselves to first order
in $U$, i.e., we apply $H_U$ only once in the
series~(\ref{BW_pert}). We then find the diagrams shown in
Fig.~\ref{Singlet}{\em c}. As the diagrams are time ordered,
$I^{(1)}_A$ differs from $I^{(1)}_B$, etc. Also, applying $H_U$
over a doubly wiggled line deserves some comment (see, e.g.,
$I^{(1)}_B$): Such a diagram would not be unambiguous since it is
not clear whether $H_U$ acts while the $f$-level is empty or
occupied, when we expand the renormalized empty $f$-propagator as
in Fig.~\ref{renorm}. For that reason we define that $H_U$ acts
while the (bare) $f$-level is empty. The other case yields a
different diagram (here, $I^{(1)}_C$). The contributions of the
diagrams of Fig.~\ref{Singlet}{\em b} are given by
\begin{eqnarray}
I^{(1)}_A(z)
& = & -\frac{2UV^4}{N_f N_s^3} \sum_{qrr'R}
G_1(z+\epsilon_q)
G_1(z+\epsilon_r+\epsilon_{r'}-\epsilon_R)
G_0(z+\epsilon_r-\epsilon_R)
G_1(z+\epsilon_r)
{}~\delta_{r'-q,R-r}
\nonumber
\\
I^{(1)}_B(z)
& = & \frac{UV^4}{N_f N_s^3} \sum_{qrQR}
G_1(z+\epsilon_r)
G_0(z+\epsilon_r-\epsilon_R)
G_0(z+\epsilon_q-\epsilon_Q)
G_1(z+\epsilon_q)
{}~\delta_{q-Q,r-R}
\nonumber
\\
I^{(1)}_C(z)
& = & \frac{UV^6}{N_f N_s^4} \sum_{qrr'QR}
G_1(z+\epsilon_r)
G_0(z+\epsilon_r-\epsilon_R)
G_1(z+\epsilon_r+\epsilon_{r'}-\epsilon_R)
\nonumber
\\
&& \qquad \times ~
G_1(z+\epsilon_q+\epsilon_{r'}-\epsilon_Q)
G_0(z+\epsilon_q-\epsilon_Q)
G_1(z+\epsilon_q)
{}~\delta_{q-Q,r-R}
\nonumber
\\
I^{(1)}_D(z)
& = & \frac{2UV^4}{N_f N_s^3} \sum_{qrQR}
G_0(z+\epsilon_q+\epsilon_r-\epsilon_Q-\epsilon_R)
G_1(z+\epsilon_q+\epsilon_r-\epsilon_R)
\nonumber
\\
&& \qquad \times ~
G_0(z+\epsilon_r-\epsilon_R)
G_1(z+\epsilon_r)
{}~\delta_{Q-q,r-R}
\nonumber
\\
I^{(1)}_E(z)
& = & \frac{2UV^6}{N_f N_s^4} \sum_{qrr'QR}
G_1(z+\epsilon_{r'})
G_1(z+\epsilon_q+\epsilon_r+\epsilon_{r'}-\epsilon_Q-\epsilon_R)
G_0(z+\epsilon_q+\epsilon_r-\epsilon_Q-\epsilon_R)
\nonumber
\\
&& \qquad \times ~
G_1(z+\epsilon_q+\epsilon_r-\epsilon_Q)
G_0(z+\epsilon_q-\epsilon_Q)
G_1(z+\epsilon_q)
{}~\delta_{Q-q,r-R} ~.
\label{sum_S11}
\end{eqnarray}

According to Eq.~(\ref{BW_pert}), the ground-state energy (relative to
$E_{{\rm FS}}$) is given by the smallest solution of
\begin{equation}
\displaystyle E_{\rm S}
= I^{(0)}(E_{\rm S})
+ I^{(1)}(E_{\rm S})
+ \sum_{i=A}^{E} I^{(1)}_i(E_{\rm S}) ~.
\label{E_sing}
\end{equation}
There is no contribution $\propto UV^0$ in this expression for
the ground-state energy since we introduced the Hubbard
interaction in normal ordered form in Eq.~(\ref{Ham_c}) and
restricted to first order in $U$. Hence the Hubbard interaction
enters only via the hybridization $V$ in the ground-state
energy. Expanding Eq.~(\ref{E_sing}) in $1/N_f$ we obtain
\begin{eqnarray}
E_{\rm S}^{\phantom{(0)}}
& = &
E_{\rm S}^{(0)} + \frac{1}{N_f} E_{\rm S}^{(1)} + o(1/N_f)^2~,
\nonumber
\\
E_{\rm S}^{(0)}
& = &
I^{(0)}(E_{\rm S}^{(0)}) \phantom{\frac{1}{N_f}}
\label{E_sing_exp}
\\
E_{\rm S}^{(1)}
& = &
\frac{I^{(1)}(E_{\rm S}^{(0)})
+ \sum_{i=A}^{E} I^{(1)}_i(E_{\rm S}^{(0)})}
{1 - \partial I^{(0)}(E_{\rm S}^{(0)}) / \partial E_{\rm S}}~.
\nonumber
\end{eqnarray}

{\em Diagrams for the multiplet.}  We now turn to the
ground-state energy of a multiplet state. To order $(1/N_f)^0$ it
is given by $E_{\rm M} = \epsilon_f$ (relative to $E_{\rm
FS}$). There is only one diagram contributing to order $1/N_f$,
which is shown in Fig.~\ref{Multiplet}. It is
\begin{equation}
\displaystyle J^{(1)}(z)
= \frac{V^2}{N_f N_s} \sum_Q R_0(z-\epsilon_Q) ~,
\label{sum_M}
\end{equation}
and we find therefore
\begin{equation}
E_{\rm M} = \epsilon_f + J^{(1)}(\epsilon_f) + o(1/N_f)^2 ~.
\label{E_mult}
\end{equation}

{\em Kondo temperature.}  We associate the Kondo temperature,
$T_{\rm K}$, with the difference between singlet and multiplet
ground-state energy\cite{Gunnarsson85} (in units of the band
width)
\begin{equation}
T_{\rm K}
= (E_{\rm M} - \mu - E_{\rm S}) \rho  ~.
\label{T_k}
\end{equation}
With this definition we find from Eqs.~(\ref{E_sing_exp}) and
(\ref{E_mult}) to order $(1/N_f)^0$
\begin{equation}
T_{\rm K}^{(0)}
= (\epsilon_f - \mu) \rho
- I^{(0)}(\epsilon_f-\mu-T_{\rm K}^{(0)}/\rho) ~.
\label{T_k_00}
\end{equation}
Assuming a constant density of states we have
\begin{eqnarray}
I^{(0)}(z)
& = & \rho V^2
\int_{-D}^\mu d\epsilon~ \frac{1}{z + \epsilon - \epsilon_f}
\nonumber
\\
& = & \rho V^2
\log \left|\frac{(\epsilon_f - \mu) - z}
{(D+\mu) + (\epsilon_f - \mu) - z} \right| ~,
\end{eqnarray}
and hence
\begin{equation}
T_{\rm K}^{(0)}
= (\epsilon_f - \mu) \rho
- (\rho V)^2 \log \frac{T_{\rm K}^{(0)}}{ n + T_{\rm K}^{(0)}} ~,
\label{T_k_0}
\end{equation}
where $n= (D+\mu) \rho$ denotes the filling per spin. This is
solved for small $J_{\rm K} = - V^2/(\epsilon_f - \mu)$ by
\begin{equation}
\displaystyle T_{\rm K}^{(0)}
= n \exp\left(-\frac{1}{\rho J_{\rm K}}\right) ~,
\label{T_kondo_0}
\end{equation}
cf.~Eq.~(\ref{TK0}). To order $1/N_f$, we find from
Eqs.~(\ref{E_sing_exp}), (\ref{E_mult}), and (\ref{T_k})
\begin{equation}
\displaystyle T_{\rm K}^{(1)} =
\rho \left[ J^{(1)}(\epsilon_f)
- \frac{I^{(1)}(E_{\rm S}^{(0)})
+ \sum_{i=A}^{E} I^{(1)}_i(E_{\rm S}^{(0)})}
{1 - \partial I^{(0)}(E_{\rm S}^{(0)}) /\partial E_{\rm S}}
\right] ~.
\label{T_kondo_1}
\end{equation}

{\em Connection to variational approach.} To find the
result~(\ref{T_kondo_1}) variationally the following states,
which occur as intermediate states in the diagrams, have to be
included in the trial state for the singlet ground state
\begin{eqnarray}
\hat f^\dagger_\sigma c^{\phantom{\dagger}}_{q\sigma}; ~
c^\dagger_{Q\sigma} c^{\phantom{\dagger}}_{q\sigma}; ~
c^\dagger_{Q\sigma} c^{\phantom{\dagger}}_{q\sigma}
\hat f^\dagger_{\sigma'} c^{\phantom{\dagger}}_{q'\sigma'}
|{\rm FS}\rangle &&
\label{var_states_01}
\\
c^\dagger_{Q\sigma} c^{\phantom{\dagger}}_{q\sigma}
c^\dagger_{Q'\sigma'} c^{\phantom{\dagger}}_{q'\sigma'}; ~
c^\dagger_{Q\sigma} c^{\phantom{\dagger}}_{q\sigma}
c^\dagger_{Q'\sigma'} c^{\phantom{\dagger}}_{q'\sigma'}
\hat f^\dagger_{\sigma''} c^{\phantom{\dagger}}_{q''\sigma''}
|{\rm FS}\rangle &&~.
\label{var_states_2}
\end{eqnarray}
The variational coefficients are determined up to first order in
$U$ and to leading order in $1/N_f$.\cite{Schork95} In the free
case ($U=0$) the first state corresponds to the ansatz of Varma
and Yafet, cf.\ Eq.~(\ref{VY_orig}), which gives the result
correctly to order $(1/N_f)^0$. The next two yield the $1/N_f$
corrections while the last two are of order $(1/N_f)^2$.

\section{Estimating the Kondo temperature}
\label{sec:estimate}

In this section, we estimate the effect of the diagrams $\propto
1/N_f$ on the Kondo temperature. We scale the energies by $\rho$
and study the dependence on $T_{\rm K}^{(0)}$ rather than on
$E_{\rm S}^{(0)}$ as $T_{\rm K}^{(0)}$ is the small quantity. The
transformed propagators read
\begin{eqnarray}
i(x)
& = &
-\frac{1}{\rho V^2} I^{(0)}(E_{\rm S}^{(0)}-x/\rho)
=
\log\left(\frac{x+n+T_{\rm K}^{(0)}}{x+T_{\rm K}^{(0)}}\right)
\nonumber \\
g_1(x)
& = &
-\frac{1}{\rho} G_1(E_{\rm S}^{(0)} - (x/\rho-\mu))
= \frac{1}{T_{\rm K}^{(0)} + x}
\nonumber \\
g_0(x)
& = &
-\rho V^2 G_0(E_{\rm S}^{(0)}-x/\rho)
\nonumber \\
& = &
\frac{\rho^2 V^2}
{x + T_{\rm K}^{(0)} - \rho(\epsilon_f-\mu) - \rho^2 V^2 i(x)}
\end{eqnarray}
and depend implicitly on $T_{\rm K}^{(0)}$ and $n$. The
empty-state propagator, $g_0(x)$ diverges $\propto T_{\rm
K}^{(0)}/x$ for $x\to 0$. [This corresponds to the spin
fluctuation peak at $z=E_{\rm S}^{(0)}$ in $G_0(z)$.] This
singularity yields contributions $\propto T_{\rm K}^{(0)}
\left(\log T_{\rm K}^{(0)}\right)^\nu$ to $I^{(1)}$, $I^{(1)}_i$,
and $J^{(1)}$, which we neglect against terms which remain
constant as $T_{\rm K}^{(0)}\to 0$. For larger $x$ however,
$g_0(x)$ drops slower than $1/x$ resulting in finite
contributions. In this intermediate $x$ range ($T_{\rm
K}^{(0)}\ll x\ll n$), we may safely approximate
\begin{equation}
g_0(x) \sim \rho J_{\rm K}
\label{g_0}
\end{equation}
for small $T_{\rm K}^{(0)}$. The validity of this replacement for
the whole $x$ range in the diagrams has been checked numerically.

{\em Diagrams of order $U^0$.} We begin with the contributions
$\propto U^0$ in Eq.~(\ref{T_kondo_1})
\begin{eqnarray}
I^{(1)}(E_{\rm S}^{(0)})
& = &  -\frac{\rho V^2}{N_f}
\int_0^n du \int_0^{1-n} dx ~ g^2_1(u) g_0(u+x)
\label{int_i1}
\\
J^{(1)}(\epsilon_f)
& = & - \frac{1}{\rho N_f}
\int_0^{1-n} dx~ g_0(x-T_{\rm K}^{(0)}) ~,
\label{int_j1}
\end{eqnarray}
where we again assumed a constant density of states. Inserting
(\ref{g_0}) we find for the multiplet energy
\begin{equation}
\displaystyle J^{(1)}(\epsilon_f)
= -\frac{J_{\rm K}}{N_f} (1-n)  ~,
\end{equation}
where the corrections are of higher order in
$1/(\epsilon_f-\mu)\rho$. For the singlet energy we use the same
approximation for $g_0$ to obtain
\begin{equation}
\displaystyle I^{(1)}(E_{\rm S}^{(0)})
= \frac{(\rho V)^2 J_{\rm K}}{N_f} (1-n)
\left(
\frac{1}{n+T_{\rm K}^{(0)}} - \frac{1}{T_{\rm K}^{(0)}}
\right) ~.
\end{equation}
Together with the denominator in Eq.~(\ref{T_kondo_1})
\begin{equation}
\displaystyle
1-\frac{\partial I^{(0)}(E_{\rm S}^{(0)})}{\partial E_{\rm S}}
= 1-(\rho V)^2
\left(
\frac{1}{n+T_{\rm K}^{(0)}} - \frac{1}{T_{\rm K}^{(0)}}
\right)
\end{equation}
and neglecting the $1$, we find that both contributions $\propto
U^0$ cancel. (Loosely speaking, these terms describe the energy
gain due to hybridization with unoccupied states which is the
same for multiplet and singlet state.)

{\em Diagrams of order $U$.} We continue with the estimation of
the diagrams $\propto U$. The numerical evaluation of the sums
$I^{(1)}_A,\dots, I^{(1)}_E$ is difficult because of the
$\delta$-functions, which ensure momentum conservation in the
Hubbard interaction. Since we are interested only in the
qualitative behavior, we may neglect them. This implies that the
interaction $U$ acts only at the lattice site $0$ with which the
impurity hybridizes and corresponds to taking the limit of
infinite dimensions.\cite{VollhardtEmeryBook} Then the sums
$\propto U$ in Eq.~(\ref{sum_S11}) read
\begin{eqnarray}
I^{(1)}_A(E_{\rm S}^{(0)})
& = & \frac{2UV^2\rho^2}{N_f}
\log\left( \frac{T_{\rm K}^{(0)}}{n+T_{\rm K}^{(0)}} \right)
\int_{0}^{n} du \int_{0}^{1-n} dx ~ g_1(u)~g_0(u+x)~i(u+x)
\nonumber
\\
I^{(1)}_B(E_{\rm S}^{(0)})
& = & \frac{U}{N_f}
\left[
\int_{0}^{n} du \int_{0}^{1-n} dx ~g_1(u)~g_0(u+x)
\right]^2
\nonumber
\\
I^{(1)}_C(E_{\rm S}^{(0)})
& = & - \frac{UV^2\rho^2}{N_f}
\int_{0}^{n} du~dv \int_{0}^{1-n} dx~dy~
g_1(u)~g_0(u+x)~g_1(v)~g_0(v+y)
\nonumber
\\
&&  \qquad \times
\frac{i(u+x) - i(v+y)}{u+x-(v+y)}
\nonumber
\\
I^{(1)}_D(E_{\rm S}^{(0)})
& = &  \frac{2U}{N_f}
\int_0^n du~dv \int_0^{1-n} dx~dy~
g_1(u)~g_0(u+x)~g_1(u+v+x)~g_0(u+v+x+y)
\nonumber
\\
I^{(1)}_E(E_{\rm S}^{(0)})
& = &  -\frac{2UV^2\rho^2}{N_f}
\int_0^n du~dv \int_0^{1-n} dx~dy~
g_1(u)~g_0(u+x)~g_1(u+v+x)
\nonumber
\\
&&\qquad \times ~ g_0(u+v+x+y) ~
\frac{i(u+v+x+y) - i(0)}{u+v+x+y} ~.
\label{integral_u1}
\end{eqnarray}
For the integrals $I^{(1)}_A$, $I^{(1)}_B$ and $I^{(1)}_D$ it
numerically proves sufficient to replace $g(x)$ as in
Eq.~(\ref{g_0}) in the limit $T_{\rm K}^{(0)}\to 0$. Then the
following leading behavior for $T_{\rm K}^{(0)}\to0$ can be
calculated analytically
\begin{eqnarray}
I^{(1)}_A(E_{\rm S}^{(0)})
& = & \frac{UV^2\rho^2C(n)}{N_f}
\log\left( \frac{T_{\rm K}^{(0)}}{n} \right)
\nonumber
\\
I^{(1)}_B(E_{\rm S}^{(0)})
& = & \frac{U}{N_f} (1-n)^2
\label{IABD}
\\
I^{(1)}_D(E_{\rm S}^{(0)})
& = & -\frac{UC(n)}{N_f} (1-n)
\frac{1}{\log T_{\rm K}^{(0)}} \to 0
\nonumber
\end{eqnarray}
with $C(n) = -2\left[ n\log n +(1-n) \log(1-n) \right] > 0$. The
integral $I^{(1)}_E$ simplifies as one integration can be
performed:
\begin{eqnarray}
I^{(1)}_E(E_{\rm S}^{(0)})
& = & -\frac{2UV^2\rho^2}{N_f}
\int_0^n du~dv \int_0^{1-n} dx~
g_1(u)~\frac{1}{i(u+x)-i(0)}
\nonumber
\\
&& \qquad\times
g_1(u+v+x)~\log\left(\frac{u+v+x+1-n}{u+v+x}\right)~.
\end{eqnarray}
By numerical evaluation, one finds that $I^{(1)}_C$ and
$I^{(1)}_E$ remain finite as $T_{\rm K}^{(0)}\to 0$. They are,
however, small compared to $I_B^{(1)}$ as they have an additional
prefactor $\rho^2 V^2$. Keeping only diagrams $I_A^{(1)}$ and
$I_B^{(1)}$ in Eq.~(\ref{T_kondo_1}), which corresponds to
considering only the states~(\ref{var_states_01}) in a
variational calculation, we obtain in the limit of small $T_{\rm
K}^{(0)}$
\begin{eqnarray}
T_{\rm K}^{(1)}
& = &
- \frac{T_{\rm K}^{(0)}}{N_f}
\left(
\rho U C(n) \log\left( \frac{T_{\rm K}^{(0)}}{n} \right)
+ \frac{U}{\rho V^2} (1-n)^2
\right)
\nonumber
\\
& = & \frac{T_{\rm K}^{(0)}}{N_f}
\left(
\frac{U}{J_{\rm K}} C(n)
- \frac{U}{\rho V^2} (1-n)^2
\right) ~.
\label{res_fin}
\end{eqnarray}
The first contribution to $T^{(1)}_{\rm K}$ is positive and,
therefore, enhances the Kondo temperature. Since it depends on
$U/J_{\rm K}$ it is related to spin degrees of freedom. A similar
contribution was found in Ref.~\onlinecite{Khaliullin94} and it
was attributed to the enhancement of spin fluctuations that
result from the reduction of charge degrees of freedom when
turning on $U$. The second contribution in Eq.~(\ref{res_fin})
depends on $\rho V^2$ rather than $J_{\rm K}$. It is related with
charge degrees of freedom. A similar effect has been found in
Ref.~\onlinecite{Schork94} and has been interpreted as the
increase in energy of the virtual state in the spin-exchange
process because in the virtual state a conduction site is doubly
occupied. It decreases the Kondo temperature. However, in the
limit $ \rho |\epsilon_f-\mu|\gg 1$ that we considered
throughout, the first term dominates: Overall we find an increase
of the Kondo temperature.

This interpretation can be put onto more solid grounds by the
following observation: If we would not scale the Hubbard
interaction among the conduction electrons by $1/N_f$, the
corrections due to $U$, $I^{(1)}_i(z)$, would be of the same
order as $I^{(0)}$. Then the integral $I^{(1)}_B(z)$, which
remains constant as $z\to 0$, would effectively shift the
position of the $f$-level to $\epsilon_f^*$, whereas
$I^{(1)}_A(z)\sim \log z$ would renormalize the exchange coupling
$V^2/\epsilon_f^*$ [cf.\ Eqs.~(\ref{T_k_00}) and
(\ref{T_k_0})]. However, without scaling there would be
contributions of higher order in $U$ which diverge as
$N_f\to\infty$.

{\em Comparison to previous results.} In
Ref.~\onlinecite{Khaliullin94}, a Kondo model ($N_f=2$) with
correlated conduction electrons has been investigated to lowest
order in the interaction strength $\tilde U$ by a mean-field
decoupling of the Kondo-exchange interaction. The following
increase of the Kondo temperature has been found
\begin{equation}
\displaystyle \frac{T_{\rm K}(\tilde U)}{T_{\rm K}(0)}
= \exp\left(\frac{\alpha}{\rho J_{\rm K} (1+\alpha)} \right)
\label{Kha}
\end{equation}
with $\alpha = (3/2) \rho \tilde U \log 2$. In contrast to our
result~(\ref{res_fin}), the increase of the Kondo temperature
seems to depend exponentially on $\tilde U$. Note, however, that
in our treatment the interaction had to be scaled by
$1/N_f$. Scaling $\tilde U$ in Eq.~(\ref{Kha}) and expanding in
$1/N_f$ yields the first term of our result~(\ref{res_fin}) (with
a factor $\log 2$ for $N_f = 2$ at half filling instead of $3/4
\log 2$).

The second term of Eq.~(\ref{res_fin}), which describes the
effective shift of the $f$-level, cannot be found in
Ref.~\onlinecite{Khaliullin94} because there the Kondo model has
been investigated, where the charge degrees of freedom of the
impurity have already been projected out. Therefore, $J_{\rm K}$
in~(\ref{Kha}) is an effective coupling constant which depends on
$U$.\cite{Schork94}

Without scaling the Hubbard interaction, all integrals would be
of order $(1/N_f)^0$. As discussed above, the position of the
$f$-level and the Kondo coupling constant are modified, and these
corrections occur in the exponent as in Eq.~(\ref{Kha}).

\section{Conclusion}
\label{sec:conclusion}

The aim of this paper was to investigate the influence of
correlations among the conduction electrons on the Kondo
effect. Lead by the situation prevailing in
Nd$_{2-x}$Ce$_x$CuO$_4$, we proposed a model with a twofold
degenerate impurity ($N_f=2$) which hybridizes with a correlated
$s$-band and straightforwardly generalized it to arbitrary
$N_f$. As discussed in the introduction, this generalization does
not correspond to the physical situation of an $N_f$-fold
degerate impurity hybridizing with a correlated $s$-band for
$N_f>2$, in contrast to the uncorrelated case. Although
artificial, we saw that this model allows for systematically
studying the effects of the correlations on those diagrams which
are usually considered in the uncorrelated case.

In particular, we assumed that the correlations are weak and
calculated their effect on the Kondo temperature to lowest order
in $1/N_f$. We found two competing effects: The first
contribution is related to charge fluctuations. Because the
energy of the virtual state in the spin exchange process
increases, the Kondo temperature is reduced. This corresponds
effectively to a shift of the position of the $f$-level. A
similar effect has been found in Ref.~\onlinecite{Schork94}. The
second contribution is related to the enhancement of spin
fluctuations of the conduction electrons. The Kondo exchange
coupling is effectively enhanced and the Kondo temperature
increases. In the Kondo limit the second contribution dominates
the first one, so that we find in total an increase of the Kondo
temperature for small $U$.

To our opinion more interesting is that corrections to the Kondo
temperature occur already to lowest order in the Hubbard
interaction $U$. To obtain them in a variational approach, trial
states are needed which in the uncorrelated case yield
corrections to the ground-state energy which are of order $1/N_f$
(and higher). Thus, our result cannot be obtained by an ansatz of
the Varma-Yafet type~(\ref{VY_orig}). This shows that the effect
of the Hubbard correlations is more intricate than just to
provide quasiparticles with a modified density of states at the
Fermi surface which hybridize with the $f$-orbital as it was
described by Varma and Yafet for the uncorrelated case, $U=0$.

If we wish to proceed to higher order in $U$, we note that to
order $1/N_f$ only RPA-type diagrams contribute since each $U$
vertex carries a factor $1/N_f$, which has to be compensated by a
spin summation, i.e., a closed loop of conduction electrons (in
fact, the ground-state energy of the Hubbard model~(\ref{Ham_c})
to order $1/N_f$ is given by summing the diagrams of RPA type and
neglecting the $\sigma\neq\sigma'$ constraint). Only few more
intermediate states will occur. This is, however, an artifact of
our scaling of the Hubbard interaction and one expects that in a
realistic model (without the restrictive scaling of $H_U$ and
finite $N_f$) intermediate states with more and more excited
electron-hole pairs contribute, the number of which increases
with increasing order of $U$. Therefore, it seems questionable
that a systematic $1/N_f$ treatment grasps the correct physics
for realistic models of interacting conduction electrons in the
limit of strong correlations.

\acknowledgements

The author has benefited from numerous discussions with Professor
P.~Fulde, Dr.~G.~Khaliullin, and Dipl.-Phys.~K.~Fischer.

\begin{figure}
\caption{Renormalization of the empty state propagator to order
$(1/N_f)^0$}
\label{renorm}
\end{figure}

\begin{figure}
\caption{Diagrams for the singlet ground-state energy.
{\em a.}~Order $U^0$ and $(1/N_f)^0$, $I^{(0)}(z)$.
{\em b.}~Order $U^0$ and $(1/N_f)^1$, $I^{(1)}(z)$.
{\em c.}~Order $U^1$ and $(1/N_f)^1$,
$I_i^{(1)}(z)~(i=A,\dots,E)$}
\label{Singlet}
\end{figure}

\begin{figure}
\caption{Diagram for the multiplet ground-state energy,
$J^{(1)}(z)$}
\label{Multiplet}
\end{figure}

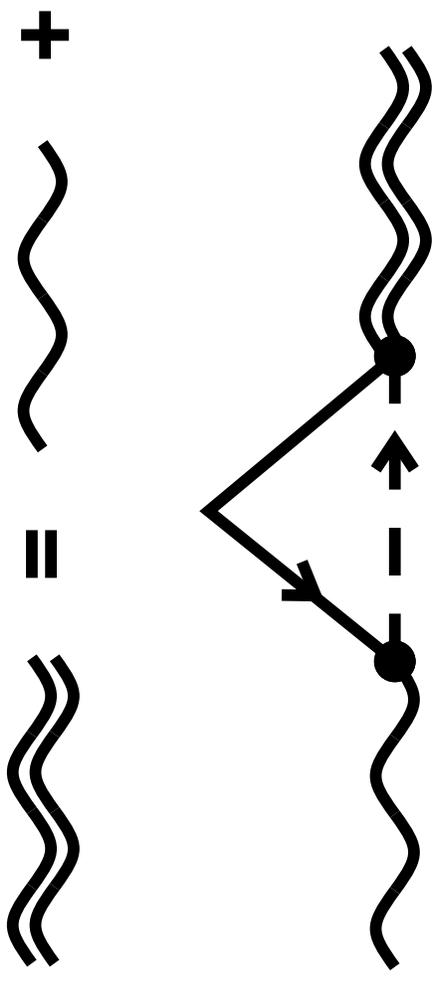
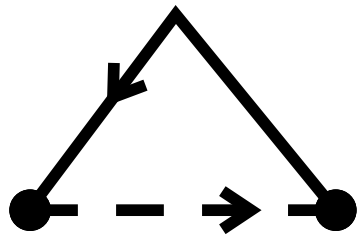
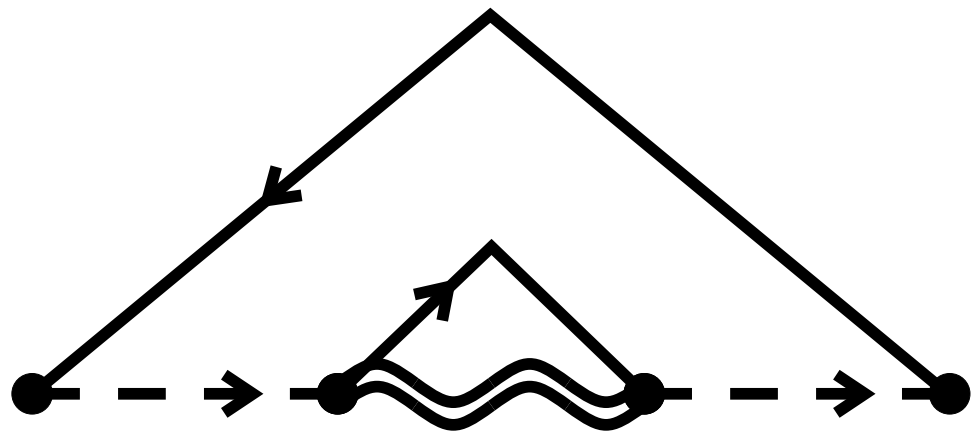
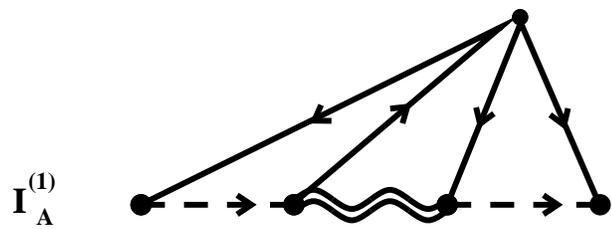
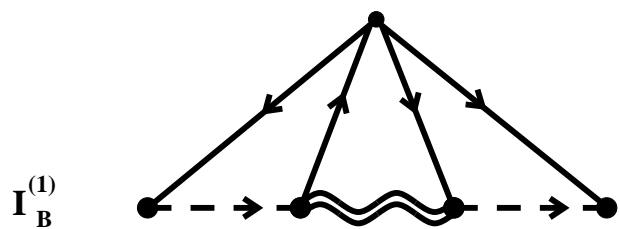
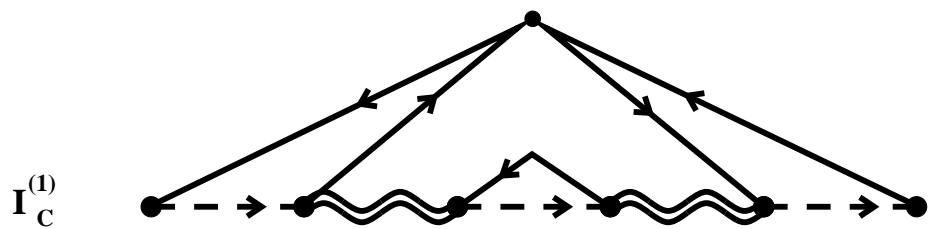
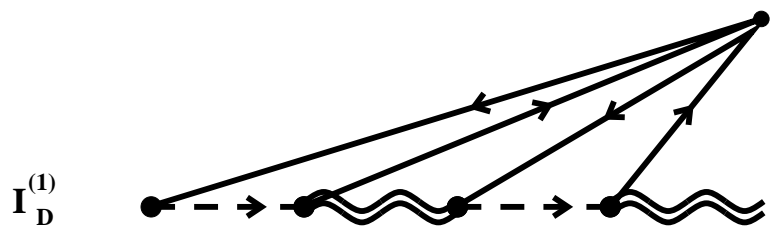
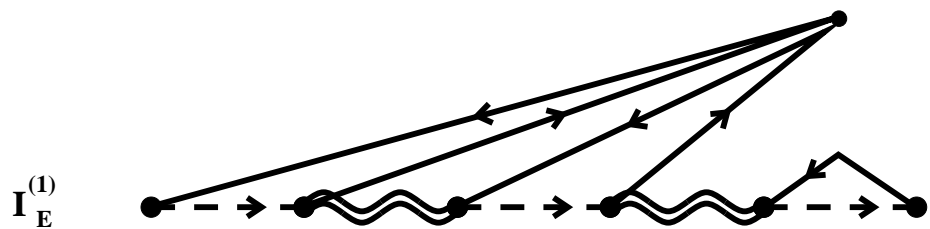
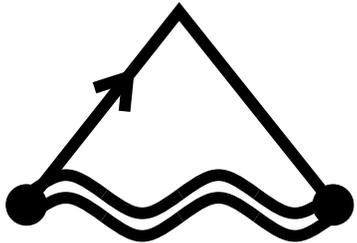
\end{document}